\documentstyle[12pt,aasms4]{article}
\begin{document}
\title{ THE $^7Be$ ELECTRON CAPTURE RATE IN THE SUN}
\author{Andrei V. Gruzinov \& John N. Bahcall}
\affil{Institute for Advanced Study, School of Natural Sciences,
Princeton, NJ 08540}

\begin{abstract}
For solar conditions, we numerically integrate the density matrix equation for 
a 
thermal electron in the field of a $^7Be$ ion and other plasma ions and 
smeared-out electrons. Our results are in agreement with previous calculations 
that are based on a different physical picture, a picture which postulates the existence of distinct continuum and 
bound state orbits for electrons. The density matrix calculation of the electron capture rate is independent of the nature of electron states in the solar plasma. To within a 1\%  accuracy, the effects of screening can 
be described at high temperatures by a Salpeter-like factor of 
$\exp (-Ze^2/kTR_D)$, which can be derived from the density matrix equation. The total theoretical uncertainty in the electron capture rate 
is about $\pm 2\%$.
\end{abstract}
\eject

\section{Introduction}
\label{int}

Observations of solar neutrinos by  four different experiments (Davis 1994; Hirata et al. 1991; Anselmann et al. 1995; Abdurashitov et al. 1994) have 
revealed important information about the interior of the sun and also about 
neutrino properties (Bahcall 1996).  Further experiments are underway to study in more 
detail the rare high-energy neutrinos from $^8$B beta-decay (Takita 1993; McDonald 1994; Icarus Collaboration 1995) and the lower
energy neutrinos from the relatively common  $^7$Be electron 
capture in the sun (Arpesella 1992).  The solar rate of electron capture on the ambient 
$^7$Be ions is almost a thousand times faster than the rate of proton-capture 
(which produces
$^8$B neutrinos).  Therefore, the predicted flux of  $^8$B neutrinos that is 
studied in the Kamiokande, Superkaniokande, SNO, and ICARUS experiments is 
inversely proportional to the electron capture rate on $^7$Be.  

Over the past 35 years, a number of different studies have been undertaken 
(Bahcall 1962; Iben, Kalata, \& Schwartz 1967; Bahcall \& Moeller 1969; Watson \& Salpeter 
1973;  Johnson, Kolbe, Koonin, \& Langanke 1992)  to calculate accurately the 
rate at which $^7$Be ions in the solar 
plasma capture electrons from continuum and bound states.  Successive 
improvements have been introduced into the calculations, but in all cases the 
changes have been rather small.

All previous calculations have been based upon a simplified model of 
the solar plasma in which the different quantum configurations were idealized 
into separate bound and continuum states. Bound electron captures were imagined to occur from isolated atoms in which the plasma was represented by a mean field (Iben et al. 1967).

In the present treatment, we evaluate directly
 the total electron capture rate on 
$^7$Be ions by integrating numerically the density matrix equation for a 
thermal electron in the field of a $^7$Be nucleus in  plasma environment. We use Monte Carlo simulations 
to represent the relatively large fluctuations that result from the small 
number of ions within a Debye sphere. Our technique avoids  the 
necessity for defining separate bound states within the solar plasma and 
allows us to take account of the departures from spherical symmetry that 
result from the 
fluctuations in the number of ions near the $^7$Be nucleus. We thus finesse complicated questions concerning the properties and meaning of electron ``bound states'' in a dense plasma in which electrons have appreciable probabilities to be at different ion sites (cf. discussion of the Stark effect in \S 3 of the paper)

In \S 2, we summarize previous results obtained with the mean field 
approximation.  We then give in \S 3 an approximate analytic calculation 
that suggests that  the effects of fluctuating electric fields caused by ions 
near the $^7$Be nucleus might change significantly the capture rate for bound 
electrons.  We summarize the density matrix formulation in \S 4 and 
present the results of our Monte Carlo simulations in \S 5. We provide in \S 6 a heuristic argument why, as found in numerical simulations, the effects of fluctuations on a total electron capture rate are small.  We summarize our results in \S 7.

\section {Mean Field Screening}
\label{mean}

The rate of electron capture is proportional to the density of electrons at 
the 
$^7Be$ nucleus. In the solar plasma, there are continuum (positive energy) and 
bound (negative energy) electrons. If the plasma density is sufficiently low, 
the density of continuum electrons at the nucleus is greater than the mean 
plasma density by a well-known Coulomb correction factor (Bahcall 1962)
\begin{equation}
w_c~=~<|\psi _k(0)|^2>~=~<{2\pi /k\over 1-e^{-2\pi /k}}>.
\end{equation}
For the non-relativistic solar plasma, $k=v/Z$ is the wavenumber, and the 
brackets indicate the average over thermal distribution of electron 
velocities,  
$v$. Atomic units ($\hbar =e=m_e=1$) are used here, and throughout the paper. The $^7Be$ electron 
capture is maximal at $R/R_{\odot }=0.06$ (Bahcall 1989), where the inverse 
temperature is $\beta =0.0215$ (Bahcall \& Pinsonneault 1995). For this 
typical 
solar environment, the density enhancement at the nucleus due to  
electrons in continuum states is $w_c=3.18$. 

\subsection {Bound States plus Continuum States}

Iben, Kalata, \& Schwartz (1967) pointed out that under solar conditions bound 
electrons give a substantial contribution to the density at the nucleus. The 
bound state enhancement factor is given by
\begin{equation}
w_b=\pi ^{1/2}(2\beta )^{3/2} \sum n^{-3}\exp (\beta Z^2/2n^2),
\end{equation}
and equals $w_b=1.20+0.21=1.41$, where $w_{b1}=1.20$ is the ground state 
contribution. The total density enhancement factor is $4.59$. 

Iben, Kalata, \& Schwartz (1967) realized that Debye-H\"uckel screening would 
reduce electron densities at the nucleus for bound electrons and evaluated 
this 
reduction for isolated atoms. We first present the results of the Iben et al. model.

Table 1 gives the calculated ground state 
ionization 
potentials, $\chi$, and the probability densities, $\psi ^2$, at the nucleus
 for a 
screened Coulomb potential with $Z$ taking on values from 1 to 6 and Debye radius $R_D=0.45$ , 
which is the solar value at $R/R_{\odot }=0.06$. For $Z=1$, Debye-H\"uckel 
screening destroys all the bound states. Following Iben et al., we define the rate reduction 
factors, $F_{IKS}$, by  which the bound state capture rate is reduced due to 
screening,
\begin{equation}
F_{IKS}=\psi^2e^{\beta \chi}/\psi_0^2e^{\beta \chi_0},
\end{equation}
where the subscript $0$ indicates unscreened values. Thus, we see from Table 1 that bound state 
screening 
reduces the total capture rate by a factor 
\begin{equation}
R=(w_c+F_{IKS}w_{b1})/(w_c+w_b)=0.85, 
\end{equation}
or by 15\% .

Screening effects on continuum electrons were studied by Bahcall \& Moeller 
(1969), who integrated numerically the Schroedinger equation for continuum electrons. For $^7Be$ under solar conditions, screening corrections are small but 
larger 
than our calculational accuracy. Let the screening corrections for continuum 
electrons be represented by
\begin{equation}
F_{BM}=<\psi ^2>/<\psi _0^2>.
\end{equation}
Table 1 gives values of $F_{IKS}$ and $F_{BM}$ for different nuclear charges
 $Z$; solar values at $R/R_{\odot }=0.06$
 were used for $\beta$ and $R_D$. 

The total electron capture rate should 
be 
calculated using a density enhancement factor
\begin{equation}
w_{IKSBM}=F_{BM}w_c+F_{IKS}w_{b1},
\end{equation}
where we make the excellent approximation that screened excited bound states give a negligible 
contribution. For $Z=4$, Eq. (6) gives 
$w=0.978\times 3.18+0.62\times 1.20=3.85$, 
 which is 16\% smaller than the unscreened value of 4.59. 

\subsection {Salpeter Formula}

The numerical results summarized by Eq. (6) are well approximated by a simple analytical expression analogous to the formula derived by Salpeter 
(1954) for weak screening of thermonuclear reactions. The derivation is 
simple. Consider a screened potential in the vicinity of the origin, $r=0$. 
The first order expansion of the potential gives
\begin{equation}
\phi ={Z\over r}e^{-r/R_D}\approx {Z\over r}-{Z\over R_D}.
\end{equation}
Thus the potential near the nucleus is a Coulomb potential plus an approximately constant correction. In 
statistical equilibrium, the constant change in the potential reduces the 
electron density at the nucleus by a Boltzmann factor, $F_S=\exp (-\beta 
Z/R_D)$, and the density enhancement factor is given by
\begin{equation}
w_S=F_S(w_c+w_b). 
\end{equation}
Table 1 compares, in the last two rows, our numerical values obtained from the detailed quantum mechanical calculations summarized by Eq. (6), 
and the simple Salpeter-like formula, Eq. (8). The agreement between the two results is about 1\% for 
$Z$ 
less than 6.

\section{Fluctuations and the Naive Stark Effect}
\label{stark}

The density enhancement obtained previously by solving the Schroedinger 
equation, Eq. (6), or by the statistical equilibrium argument, Eq. (8), is based 
on a model which represents the solar plasma by a screened nucleus and a sea 
of 
non-interacting electrons. At $R/R_{\odot }=0.06$, the effective plasma density is 
$n=(8\pi \beta R_D^2)^{-1}=9.1$, and there are on average only 3.5 ions in a Debye sphere (we use a hydrogen plasma model). Watson \& Salpeter 
(1973) suggested that the small number of ions in the Debye sphere 
implies 
that thermal fluctuations in the screening might be of importance. They 
calculated corrections due to a fluctuating number of ions close to the 
nucleus. 
Spherical symmetry was assumed in their calculations, that is, plasma ions were represented by spherical 
shells centered at the nucleus. Watson \& Salpeter found a 7\% decrease in the 
bound state capture rate, which implies a 2\% decrease in a total capture rate.

Asymmetric fluctuations might plausibly  have an even stronger effect on the 
bound state capture rate. We give a crude argument that shows that the effects 
of fluctuations must be evaluated carefully. In the first approximation, 
asymmetry implies that an ion (e.g., a $^7Be$ nucleus) experiences an electric field,  ${\cal E}$, 
produced by the other plasma ions and smeared-out plasma electrons. The electric 
field 
changes the ground state energy $-\chi$ (the Stark effect) and the probability 
density at the nucleus $\psi ^2$. The first effect increases the capture rate 
since the ionization potential increases. In second order of the perturbation 
theory (eg. Landau \& Lifshitz 1977)
\begin{equation}
\chi ={Z^2\over 2} +{9{\cal E} ^2\over 4Z^4}.
\end{equation}
On the other hand, the value of the wave-function at the nucleus decreases,
\begin{equation}
\psi =2-{81{\cal E} ^2 \over 8Z^6},
\end{equation}
which reduces the capture rate. The two effects together alter the ground 
state 
electron density by a factor, $F_{\cal E}$, where
\begin{equation}
F_{\cal E}={\psi ^2\over 4}e^{\beta \delta \chi}\approx 1-(81/8-9\beta 
Z^2/4){{\cal E} ^2 \over Z^6} \approx 1-.0023{\cal E} ^2,
\end{equation}
for $Z=4$ and $\beta =0.0215$. Thus, the fluctuating electric field suppresses 
the bound state capture rate.

 To estimate quantitatively the reduction factor due to the Stark effect, we 
need to know the size of the fluctuating field, ${\cal E}^2$. For an 
illustrative model calculation, one can use the well-known Holtsmark 
probability 
distribution for the electric field. This distribution is exact for unscreened 
non-interacting ions (low densities). The Holtsmark probability distribution is
\begin{equation}
P_H(x)={2x\over \pi}\int\limits_0^\infty dy\sin (xy)y\exp (-y^{3/2}).
\end{equation}
Here $x$ is the normalized electric field, $x={\cal E} /{\cal E}_H$, and the 
characteristic electric field is 
\begin{equation}
{\cal E}_H=2.6n^{2/3}\approx 11.
\end{equation}
We do not consider very large electric fields, $x\gg 1$, since in this domain 
the assumption of non-interacting ions is obviously incorrect. We therefore 
take 
$P(x)=0$ for large $x$ where the factor (11) becomes negative. We find, upon averaging Eq. (11), 
$F=0.21$ 
- a strong effect.

Both second order perturbation theory and the Holstmark distribution were used 
in the calculation outside of their domains of applicability. Also the effects of the fluctuating 
fields on the continuum capture rate were not considered. We do see from these simplified arguments that fluctuation 
effects 
on the electron capture rate have to be carefully investigated.

\section {Density Matrix Formulation}
\label{matrix}

To compute the mean density of electrons at a nucleus, one could solve the 
Schroedinger equation many times for a large, representative set of 
distributions of ions and smeared electrons. Given the numerical results for a 
set of configurations, the density at the nucleus would be computed as the 
average of the individual densities over the Boltzmann weighted ion 
configurations. This is a difficult but, fortunately, unnecessary task. We do not 
even have to know all the quantum states for a given ion configuration, which 
is 
a hard problem by itself.

For a given ion configuration, we are interested in just one number - the 
density of thermal electrons at the nucleus. The average density can be 
calculated by solving the density matrix equation (e.g. Feynman 1990)
\begin{equation}
\partial _\tau \rho =\{ {1\over 2}\nabla ^2+{Z\over r}e^{-r/R_D}+V(r)\} \rho, 
\end{equation}
\begin{equation}
\rho (r,\tau =0)=\delta ^{(3)}(r).
\end{equation}
Here $V$ is the fluctuating potential created by neighboring ions and smeared 
electrons. In the density matrix formulation, bound and continuum electrons 
are 
treated equally. One solves Eq. (14) for a number of different 
realizations 
of $V$ and computes the average $<\rho _Z(r=0,\tau =\beta )>$. The density enhancement factor, $w$, is then given by the ratio
\begin{equation}
w={<\rho _Z(0,\beta )>\over \rho _0(\beta )},
\end{equation}
where $\rho _0(\beta )$ is the normalization factor computed by averaging the solution of Eq. (14) with $Z=0$ (for $V=0$, one has $\rho _0(\beta )=(2\pi \beta )^{-3/2}$).

Equation (14) makes it clear why the effects of screening should be 
accurately 
described by a Salpeter-like factor $\exp (-\beta Z/R_D)$ if the temperature 
is 
high enough and the effects of the fluctuating potential are unimportant. At 
small $\beta$, the diffusing particle described by (14) stays close to the 
origin. At small distances, the expansion of the screened potential, Eq. (7), 
is 
valid. According to Eq. (14), a constant potential $U$ causes the density to be 
multiplied by a factor $\exp (U\tau )$. 

The diffusion with multiplication problem, Eq. (14), can be solved easily by 
direct three-dimensional numerical simulations for solar conditions, because 
the 
inverse temperature $\beta$ is small ($\sim 0.02$), and the diffusive 
trajectory 
stays close to the origin. We simulated Eq. (14) using a $30^3$ mesh in a cube 
with a side 0.6. This gives a spatial resolution $\Delta =0.02$, which should 
suffice because the Bohr radius for $Z=4$ is 0.25. The Coulomb potential was 
regularized by the prescription
\begin{equation}
1/r \rightarrow (r^2+\Delta ^2/7.7)^{-1/2},
\end{equation}
in all of our calculations.   

To test our code we calculated  the mean field 
theoretical results of \S 2 using our solutions of the density matrix equation. In the absence of the fluctuation potential, the 
denominator in Eq. (16) is $(2\pi \beta )^{-3/2}$. We used the code to compute 
the 
numerator of Eq. (16) for $R_D=\infty $, and for $R_D=0.45$, for $Z$ from 1 to 6. In all 
cases, our code reproduced the mean field results with an accuracy better 
than 
1\%; expression (6) was used to determine a theoretical mean field value value in the screened case.

\section {Monte Carlo Simulations of Fluctuating Fields}
\label{carlo}

We studied by Monte Carlo techniques the effects of fluctuations, $V$, on the 
density enhancement factor, $w$, for $Z=4$, $\beta =0.0215$, and the ion 
density 
$n=9.1$. Since the probability functional for the field $V$ is unknown, we 
simulated two extreme cases: randomly distributed ions (case I) and Boltzmann distributed ions (case II).

For case I, screened ions with mean density 9.1 
were randomly distributed within a cube of unit length around the $^7Be$ nucleus of 
charge $Z=4$. For case II, both the surrounding ions and the central nucleus 
were screened only by electrons ($R_D\rightarrow R_D'=\sqrt2 R_D$) but the ion 
configurations were weighted by $e^{-\beta U}$, where
\begin{equation}
U=Z\sum {1\over r_j}e^{-r_j/R_D'}+\sum {1\over r_{jk}}e^{-r_{jk}/R_D'}.
\end{equation}
Here $r_j$ are positions of ions, which were assumed to be confined to a sphere of radius 
$R=1$ around the nucleus; $r_{jk}$ are the inter-ion distances. The Boltzmann 
weights take account of the ion part of the screening. In fact, plasma ions 
that 
are at distances greater than R also contribute to the screening of the 
nucleus. 
Their contribution to the potential at the nucleus is 
\begin{equation}
\delta \phi ={Z\over 2R_D}e^{-R/R_D}.
\end{equation}
We take this additional potential into account by subtracting a small Salpeter-like correction, 
$\delta w=w\beta \delta \phi=0.04$ from the density enhancement given by 
Eq. (16). 

The probability distribution for potentials is different in the two models. In case I, the nucleus can experience arbitrarily high electric 
fields, while in case II, the stronger fields do not contribute because they have 
smaller Boltzmann weights.

Numerical results are shown in Table 2 for different values, $N_{MC}$, of 
Monte 
Carlo realizations of ion configurations. The percent deviations shown are 
fractional differences with respect to the mean field theoretical result of 
3.85. As can be seen seen from the Table, the average effects are smaller than 
1\% . We repeated the calculation for different parameters (solar 
center and the outer edge of the reaction, $R/R_{\odot }=0.15$) and got similar results - fluctuation effects change the reaction rate by less than 1\% .

\section {Heuristic Estimate of Fluctuation Effects}

Our numerical simulations show that fluctuations have little effect on the electron capture rate. However, the second order perturbative calculations in \S 3 predict a strong effect for the bound state captures. Moreover, Watson \& Salpeter (1973) suggested that the effects of fluctuations might be significant if the average number of ions in the Debye sphere, $N$, is small. In the solar case, $N$ is of the order of a few, and strong effects might be expected.

How can we understand the fact that the total capture rate is insensitive to fluctuating electric microfields? In the density matrix formulation (\S 4), electric microfields ${\cal E}$ are described by the fluctuating potential $V={\cal E}x$ in Eq. (14). As a result, both the un-normalized density at the nucleus $\rho _Z$, and the density normalization $\rho _0$ are shifted by the fluctuating field. The Coulomb attraction keeps the diffusing thermal electron (as described by the density matrix equation (14)) in the vicinity of the nucleus, and we may suppose that the shift in $\rho _Z$ is smaller than the shift in $\rho _0$. The density matrix equation for $\rho _0$ is simple, and one can calculate the shift in $\rho _0$ using the perturbation expansion of the density matrix (e.g. Feynman 1990). In second order perturbation theory, we find
\begin{equation}
{\delta \rho _0\over \rho _0}={\beta ^3 {\cal E}^2\over 24}.
\end{equation}
For the characteristic microfield (${\cal E}\sim {\cal E}_H\approx 11$) and the inverse temperature ($\beta =0.0215$), Eq. (20) gives $\delta \rho _0/\rho _0=5\times 10^{-5}$, which is indeed a small effect.

Taking the estimate (20) as an upper bound for the fluctuation effects, and substituting the Holtsmark field, Eq. (13), for ${\cal E}$, one has (restoring dimensions)
\begin{equation}
{\delta w\over w}<C{a_0\over R_D}N^{-5/3}.
\end{equation}
Here $C$ is a dimensionless number and $a_0$ is the Bohr radius. Since, $a_0\sim R_D$, Eq. (21) should give large effects when $N$ is sufficiently small. The reason for the calculated small effect of fluctuations is the size of the dimensionless number $C$, which is
\begin{equation}
C={(2.6)^2\over 24}({3\over 4\pi })^{4/3}6^{-3}=2\times 10^{-4}.
\end{equation}

\section {Summary and Discussion}

The rate of electron capture by $^7Be$ ions in the solar plasma has traditionally been computed as the sum of two different processes: capture from continuum orbits plus capture from bound orbits. But, the high density of electrons and ions in the solar interior makes this separation of quantum states a delicate issue; bound states are continually being formed and dissolved as the result of plasma interactions. We illustrate one aspect of this complexity in \S 3, where we discuss the Stark effect caused by the fluctuating electric microfield.

In this paper, we have calculated the total rate of electron capture by $^7Be$ using the density matrix formalism (cf. Eq. (14) and Feynman 1990) without reference to the individual (bound or continuum) quantum states. Our numerical code successfully reproduced the results of the previously-used mean field theory to an accuracy of better than 1\% (\S 4  and the last two rows of Table 1).

One of the most troublesome aspects of the previous mean field calculation is the possible effect of fluctuations due to the small number (about 3) of ions in a Debye sphere surrounding a $^7Be$ ion. The density matrix formulation permits us to evaluate (in \S 5) the effects of fluctuations assuming different (extreme) models for the spatial distribution of the ions. In both cases, the effect of fluctuations in the distribution of ions is less than 1\% (see Table 2).

The overall result of our calculations is to confirm to high accuracy the standard calculations for the $^7Be$ electron capture rate in the Sun (see, e.g., Iben, Kalata, \& Schwartz 1967; Bahcall \& Moeller 1969; Bahcall 1989). The results of numerical simulations (see Table 2) show that the standard formula (Bahcall 1989) is accurate to better than 1\%  .We obtained similar results for $R/R_{\odot }=0.0$, 0.06, and 0.15.

How accurate is the present theoretical capture rate, $R$? The excellent agreement between the numerical results obtained using different physical pictures (a specific model for bound and continuum states and the density matrix formulation) suggests that the theoretical capture rate is relatively accurate. The largest recognized uncertainty arises from the possible inadequacies of the Debye screening theory. Johnson et al. (1992) have performed a careful self-consistent quantum mechanical calculation of the possible effects on the $^7Be$ electron capture rate of departures from the Debye screening. They conclude that Debye screening describes the electron capture rates to within 2\% . Combining the results of Table 2 and of Johnson et al. we conclude that the total fractional uncertainty , $\delta R/R$, is small and that 
\begin{equation}
\delta R/R~<~0.02.
\end{equation}

Simple physical arguments suggest that the effects of electron screening on the total capture rate can be expressed by a Salpeter factor (see discussion in the text following Eq.(7) and Eq.(16)). The simplicity of these physical arguments provides supporting evidence that the calculated electron capture rate is robust.

\acknowledgements

We thank Dan Dubin, Freeman Dyson, and Marshall Rosenbluth for helpful 
discussions and suggestions. This work was supported by NSF PHY-9513835.
\vfil\eject

\newpage

\begin{deluxetable}{lcccccc}
\footnotesize
\tablecaption{Mean Field Theories \label{T1}} 

\tablehead{Z&1&2&3&4&5&6}
\startdata
$\chi /\chi _0$       &-    &.0034 &.12  &.25  &.35  &.43  \nl
$\psi ^2/\psi _0^2$   &-    &.098  &.52  &.70  &.80  &.85  \nl
$F_{IKS}$             &.0   &.09   &.48  &.62  &.67  &.68  \nl
$F_{BM}$              &.965 &.985  &.98  &.978 &.979 &.991 \nl
$w_{IKSBM}$           &1.38 &1.94  &2.73 &3.85 &5.50 &7.91 \nl
$w_S$                 &1.38 &1.92  &2.70 &3.81 &5.41 &7.73 \nl
\enddata

\tablecomments{The symbols represent: nuclear charge ($Z$), screened ground state ionization potential 
($\chi$) and electron density ($\psi ^2$), bound state reduction factor ($F_{IKS}$), 
continuum 
reduction factor  ($F_{BM}$), density enhancement ($w$). Results are given for an inverse temperature $\beta =0.0215$ and a Debye radius 
$R_D=0.45$.}

\end{deluxetable}

\begin{deluxetable}{lccccccc}
\footnotesize
\tablecaption{Monte Carlo Results \label{T2}} 

\tablehead{$N_{MC}$&1&2&5&10&20&50&100}

\startdata

I, $\rho _Z$  &4.84 &4.91 &4.93 &4.90 &4.99 &4.84 &4.73 \nl
I, $\rho _0$  &1.26 &1.27 &1.28 &1.27 &1.29 &1.26 &1.23 \nl
I, $w$        &3.84 &3.87 &3.85 &3.86 &3.87 &3.84 &3.85 \nl
I, \%         &-0.3 &+0.5 & 0.0 &+0.3 &+0.5 &-0.3 & 0.0 \nl
II, $\rho _Z$ &5.40 &5.46 &5.43 &5.52 &5.60 &5.57 &5.53 \nl
II, $\rho _0$ &1.37 &1.40 &1.39 &1.40 &1.44 &1.42 &1.41 \nl
II, $w$       &3.90 &3.86 &3.87 &3.90 &3.85 &3.88 &3.88 \nl
II, \%        &+1.3 &+0.3 &+0.5 &+1.3 & 0.0 &+0.8 &+0.8 \nl

\enddata

\tablecomments{The number of Monte Carlo realizations, $N_{MC}$, is given in the first row for case I. For case II, 
the number of realizations is $10\times N_{MC}$. The table also lists the average density, $\rho _Z$, the normalization ,$\rho 
_0$, and the density enhancement $w$. The last row for each case is the \% deviation from the mean field result of 3.85.}
\end{deluxetable}

\end{document}